\documentclass[vopt,a4paper,11pt]{article}
\usepackage[utf8]{inputenc}
\usepackage{amsmath}
\usepackage{amsfonts}
\usepackage{amssymb}
\usepackage{graphicx}
\usepackage{geometry}
\usepackage{textcomp}

\begin{document}

\title{On the origin of the transversal current in double layered heterostructures}
\author{ \\S.V.Iordanski\\
Landau Institute for Theoretical Physics RAS
 {142432 Russia,Chernogolovka}}

\maketitle

\begin{abstract}
It is shown that usually used theoretical model for double layered heterostructures as a pseudospin
ferromagnet \cite{1,2}, does not give the observed two dimensional spectrum. Its existence is possible only neglecting Coulomb interaction  destroying two dimensional structures and can be realized only in a strong magnetic field. That is connected also with the plain vortex lattices
arising at strong magnetic field due to thermodynamic instability \cite{4}. This model gives the reasonable
explanations of various observed effects depending on the filling of the corresponding bands.In particular in this work we show that in double layered heterostructures can exist large inter layer conductance really observed  \cite{5}.
PACS 70(73),60(67)
\end{abstract}

The experiments with double layered heterostructures have a long history beginning from the late
eighties of the previous century. The main number of theoretical works \cite{1,2} used the layer index as an additional quasi spin index. However the experimental work \cite{3} showed ,that the
theoretical results based on this assumption contradict to the performed experiment and require
some other description close to the used for one layered systems \cite{6}.

We take the coordinate $z$ perpendicular to the plain of the layers and the coordinates $\vec{r}=(x,y)$
describe a two dimensional space with a constant electron potential energy $U(z)$ (see fig.1). We
assume the existence of a constant magnetic field $B$ directed along $z$ large enough to make the energy of  electron Coulomb interaction proportional to $\sqrt{B}$ small compare to their magnetic energy
proportional to $B$.That is very important because Coulomb interaction tends to destroy two dimensional effects.Further we shall neglect Coulomb electron interaction assuming large enough    
magnetic field.

 It was shown in  the theoretical work \cite{4}  for this case the situation is unstable and the creation of plain vortex lattices give the thermodynamic gain. That is the case of the separate
action  of the variables $z$ and $\vec{r}$ therefore electron  wave functions are the products of two functions
$Z(z)\Phi(\vec{r})$. Various types of the vortex lattices are possible. The electron filling of the
proper bands define the physical properties. The separation of the variables give two Shroedinger
equations

\begin{equation}
\label{1}
i\hbar\frac{\partial Z}{\partial t}=-\frac{(\hbar)^2}{2m_e}\frac{\partial^2 Z}{{\partial z}^2}+U(z)Z(z)
\end{equation}
 where $U(z)$  is the two well potential(see fig(1).
 
 The functions $\Phi(\vec{r})$ are defined by two dimensional equation
 \begin{equation}
\label{2}
 i\hbar\frac{\partial\Phi(\vec{r})}{(\partial t)}=\hat{H}(\vec{r})\Phi(\vec{r})=\frac{1}{2m_e}\left(-i\hbar\frac{\partial}{\partial{\vec{r}}}-\frac{e}{c}\vec{A}{_{ef}}(\vec{r})\right)^2\Phi(\vec{r})
\end{equation} 
 
 Here the effective vector-potential $\vec{A}_{ef}$ include not only the vector -potential of the external magnetic field $1/2[\vec{B}\vec{r}]$ but also the contribution of vortexes with the minimal
 circulation $2\pi$ in each unit cell, energetically most favourable. The solutions of the equation (\pageref{2}) have simple band energy spectrum only at the rational values of the total magnetic flux
 through the unit cell area of the vortex lattice (see e.g. \cite{9}). Having in mind the experiments
 \cite{5} where the electron density was close to that at the half filling of Landau level, we assume
 two vortexes in each unit cell. Therefore the total magnetic flux through the area of the unit cell
 $\sigma$ will be $B\sigma-2\Phi_0=l/n\Phi_0$, where $\Phi_0=2(\pi| e|)/(c\hbar)$ is the flux quantum
 and $l,n$ are co prime integers.  We  see  that the half filling of Landau level is achieved at
 $l=0$, and the state-solutions must give the representation of Abelian group of the periodic translations with the unit cell area
 
  \begin{equation}
  \label{3}
  \sigma=\frac{2\Phi_0}{B}=\frac{4\pi |e|}{c\hbar B}
  \end{equation}
  
  The situation is close to that of graphene. It was shown in the work \cite{4} that  the lattice
  must have the hexagonal symmetry
  which is determined by the vortex lattices defining  the periodic vector-potential in the plain $\vec{r}$.
 Having in mind the experiments \cite{5}we assume that the vortex lattice has the hexagonal symmetry
 with two vortexes of the minimal circulation $2\pi$ in each elementary cell. It was shown in the work \cite{4} that in this case there are two non equivalent points $\vec{k}_0$ and $\vec{k}'_ 0$ on the boundary of the  two dimensional Billouin cell where the representation of the space group for the vortex lattice is two fold and the full filling of the lowest band corresponds to the 
 density of the half filled Landau level. We assume that the maximal energy in the lowest band corresponds to the energy $\epsilon(\vec{k_0})=\epsilon(\vec{k'}_0)$ at the critical points. We must
 add to this energy the lowest energy of the size quantization $ E_1$, corresponding to the symmetric
 wave function $Z_1(z)=Z_1(-z)$ defining the electron chemical potential
  
\begin{equation}
\label{4}
\epsilon(\vec{k}_0)+E_1=\mu> \epsilon(\vec{p})+E_1
\end{equation}
We have defined the filling only for the lowest level of size quantization.The value of the chemical potential is a common quantity for electrons  and we get for the next level $E_2$ of size quantization with antisymmetric wave function  $Z_2(z)=-Z_2(-z)$

\begin{equation}
\label{5}
\epsilon(\vec{p}')<\epsilon(\vec{k}_1)+E_1-E_2
\end{equation}
and we have two hole Fermi circles 
 $ \epsilon(p_F)=E_1-E_2$ around the critical points $\vec{k_0},\vec{k}'_0$ as is shown in fig.(2). 
 We shall consider only the case of zero temperature.

Let us consider the stationary problem without the external electric field (details can be found in
\cite{7}). Both energies $E_1$ and $E_2$ are close to the oscillation energy in one well $E_0$ without tunneling. The symmetry consideration of the tunneling process gives (in semi classical approximation) two different states with the energies $E_1$ and $E_2$, where
  
\begin{equation}
\label{6}
E_2-E_1=\frac{E_0}{\pi}\exp\left(-\frac{1}{\hbar}\int_{-a}^a 
| p | dz \right)
\end{equation}

where $-a$ and $a$  are the positions of the turning points, the integral is in the classically forbidden
region, $p=m_e v=\sqrt{2m_e(E_0-U)}$. Therefore $E_2$ and $E_1$  are exponentially close.

In the presence of the external electric field there is an additional term in the Hamiltonian

\begin{equation}
\label{7}
\delta H=-|e|\int\gamma z\hat{\rho}dz d^2 r
\end{equation}

where $\hat{\rho}$ is the quantum operator of the electron density. We choose the electric field
$\gamma>0$. Therefore we must have the current from the left well to the right one. There is no
possibility to avoid the classically forbidden region. Thus we must consider the sub barrier current. The
oscillations in the left well with the energy $E_0$ gives the in-going wave at the beginning of the
classically forbidden region 
\begin{equation}
\label{8}
\psi_{in}=\sqrt{\frac{E_0}{2\pi\hbar v}}\exp\left(\frac{i}{\hbar}\int_b^{-a}pdz-\frac{i\pi}{4}\right)
\end{equation}

where $v$ is the electron velocity. At the point $z=a$ we get the out going wave in the right well
$\psi_{out}=\psi_{in}\sqrt{D}$ where $D=\exp\left(\frac{2}{\hbar}\int_{-a}^{a}|p|dz\right)$. That
gives the electron flux to the right well (see fig. \ref{fig3}), at the turning point $z=a$
(see \cite{7})

\begin{equation}
\label{9}
\psi_{out}^+\psi_{out}v=\frac{E_0}{2\pi\hbar}\exp\left(-\frac{2}{\hbar}\int_{-a}^a |p|dz\right)
\end{equation}

Obtaining this formula we use only $Z(z)$ component of the electron wave function. For a more complete expression it is necessary to perform the second quantization using the two dimensional
representation of the space group for the vortex crystal

$$\Phi(\vec{p})\exp(i\vec{p}\vec{r}), \Phi^+(\vec{p})\exp(-i\vec{p}\vec{r})$$,
where $\Phi(\vec{p})$ is the column of two functions,  $\Phi^+(\vec{p})$ is the line,
 mutually orthogonal and normalized (see also \cite{4}). Therefore the full electron operators 
of the second quantization are
\begin{equation}
\label{10}
\psi(t,z,\vec{r})=Z(z)\sum_{\vec{p}}\exp(\frac{i\vec{p}\vec{r}}{\hbar})\Phi(\vec{p})a_1(\vec{p},t)
\end{equation}
and
\begin{equation}
\label{11}
\psi^+(t,z,\vec{r})=Z^{+}(z)\sum_{\vec{p}}\exp(\frac{-i\vec{p}\vec{r}}{\hbar})\Phi^+(\vec{p}) a_1^{+}(\vec{p},t)
\end{equation}
where $a_1(\vec{p},t), a^+_1(\vec{p},t)$ are the usual Fermi operators in Heisenberg representation
with  the energy
 $\epsilon(p)<E_1-E_2+\mu$ , here $\mu$ is the chemical potential equal to the full energy in
 Dirac points.

The quantities $Z(z) ,Z^+(z)$ do not depend on the functions $\Phi(\vec{p}),\Phi^+(\vec{p})$, but their number is essential for the counting of the different filled states. The number of the various states in the crystal band coincide with the number of the elementary cells.For the two dimensional case it
gives  $N=\frac{S}{\sigma}$,where$S$ is the sample area, and  $\sigma$ is the area of the elementary
cell. Generally speaking , we must count only the filled states and exclude the states in the hole Fermi circles  near Dirac points but their quantity  is exponential small according to eq.(\ref{6})and we neglect them.

Using eq.(\ref{9}) we get for the electron quantum transitions from the left well to the right in the unit of time

 \begin{equation}
 \label{12}
 \frac{dN_l}{dt}=\frac{S}{\sigma}\frac{E_0}{2\pi\hbar}\exp\left( -\frac{2}{\hbar}\int_{-a}^a |p|dz\right)
\end{equation}
However besides this channel of the electron transitions there is also the channel of the electron
transitions from the right well to the left with the opposite direction of the current

\begin{equation}
\label{13}
\frac{dN_r}{dt}=-\frac{S}{\sigma}\frac{E_0}{2\pi\hbar}\exp\left(\frac{2}{\hbar}\int_a^{-a}|p|dz\right)
\end{equation}
The full change of the electron number in the right well is given by a sum
$\frac{dN_t}{dt}=\frac{dN_l}{dt}+\frac{dN_r}{dt}$. In the absence of the external electric field
and $U(z)=U(-z)$ there is no current between the wells. In the presence of the electric field the
direction of the current coincide with the direction of the electric field and therefore
$$\frac{dN_t}{dt}=\frac{dN _l}{dt} $$ 
and we must calculate the sub barrier current at a small value of the electric field $\gamma$.
The effective potential energy is $ U_{eff}(z)=U(z)-\gamma |e|z $
The sub barrier imagine momentum is given by the expression
 $|p|=|\sqrt{2m_e[U_{eff}(z)-E_0]}|$ . We suppose $U(-z)=U(z)$ and a small electric field 
 $U(z)\gg|e\gamma z|$. We obtain in the linear approximation
 
 $$|p|=\sqrt{2m_e(U(z)-E_0}-\frac{\gamma|e|z \sqrt{2m_e}}{2\sqrt{U(z)-E_0}}$$
 
 Substituting this expression in the current eq.\ref{12} one must calculate
 $$\exp\left(-\frac{2}{\hbar}\int_{-a}^{a}|p|dz\right)=\exp\left(\frac{-2}{\hbar}\int_{-a}^{a}dz\sqrt{ 2m_e [U(z)-E_0}\right)\left[1+ \frac{1}{\hbar}\int_{-a}^{a}dz_1\frac{\gamma|e|z_{1}\sqrt{2m_e}}{\sqrt{U(z_1)-E_0]}}dz_1\right]$$
 The  term without $ \gamma $ can not give any contribution to the current due to its isotropy and
 must be canceled. The turning points in the linear approximation coincide with $a$ and $-a$ without
 the electric field.
  Thus we get for the current to the right well
 
\begin{equation}
\label{14}
\frac{dN_t}{d t}=\frac{S}{\sigma}\frac{E_0}{2\pi\hbar}\exp\left(
	-\frac{2}{\hbar}\int_{-a}^a dz\sqrt{2m_e[U(z)-E_0]}
\right)\frac{1}{\hbar}\int_{-a}^a dz_1\frac{|e|\sqrt{2m_e}\gamma z_1}{\sqrt{U(z_1)-E_0}}
\end{equation}

According to the accepted point of view the coefficient before a small electric field $\gamma$ 
give the conductance of the two layered heterostructure. Proportionality to the sample area is confirmed by the experiment \cite{8}. In this expression the important factor is
$\frac{S}{\sigma}=\frac{c\hbar BS}{4\pi|e|}$ which is large in a strong magnetic field and may give a large value to the product
   $ \frac{S}{\sigma}\exp\left(-\frac{2}{\hbar}(\int_{-a}^{a}dz\sqrt{2m_e[U(z)-E_0]}\right)$. 
thus explaining the high currents observed in the experiment \cite{5}. The other factors are strongly
dependent on the used semi classical approximation. The obtained result in eq.(\ref{14}) has an analog in the theory of the radioactive decay . The considered physical problem gives an example of
 the conductance due to sub barrier currents. 
 
  The vortex model is the inevitable consequence of the
 the thermodynamic instability and is connected with the magnetization in a strong magnetic field.It
 is connected with the electric field arising at the change of the magnetic field because the
 magnetic field itself can not produce the work on the electrons. Using of the vortex lattice
 model makes unnecessary the additional constructions like Chern-Simons field or the composite fermions.In the recent works \cite{10,11} authors use more complicate models assuming exact electron-hole symmetry. Another difficulty is connected with the observed electron spectrum which is far from Ll.In the vortex model the spectrum is closely connected with the separation of the variables like
 in the case of the one layered hetero structures.

Author express his gratitude  to E.Kats , I.Kolokolov and  D. Khmelnitskii for the discussions.
The work was supported by grant RSc\#14-12-08.

\begin{figure}
 \centering
 \includegraphics[width=5cm]{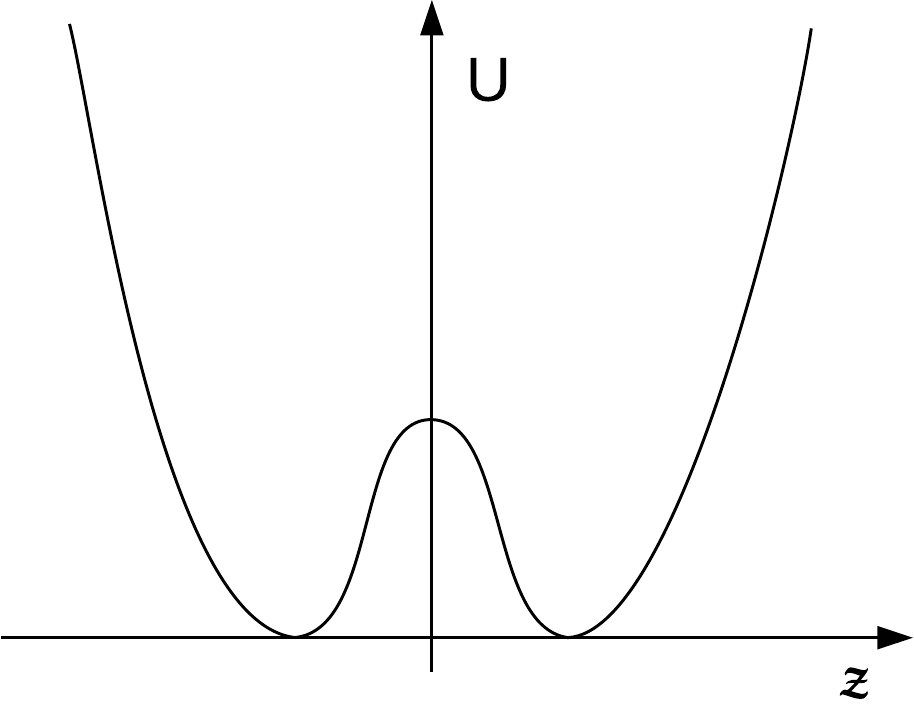}
 \caption{inter well potential}
 \label{fig1}
\end{figure}

\begin{figure}
 \centering
 \includegraphics[width=5cm]{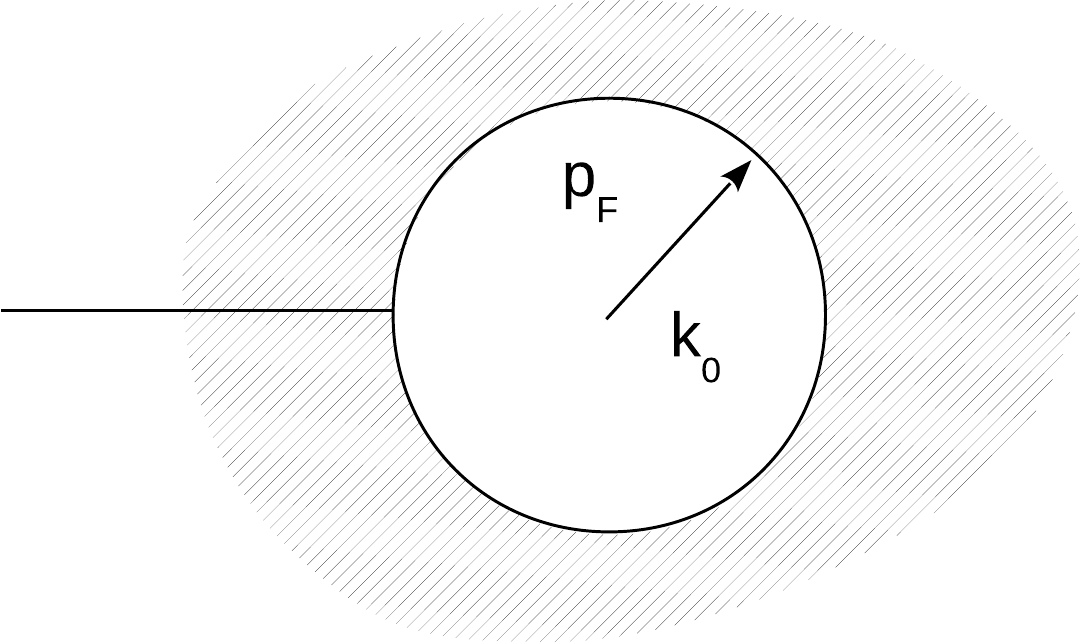}
 \caption{ the electron filling}
 \label{fig2}
\end{figure}

\begin{figure}
 \centering
 \includegraphics[width=5cm]{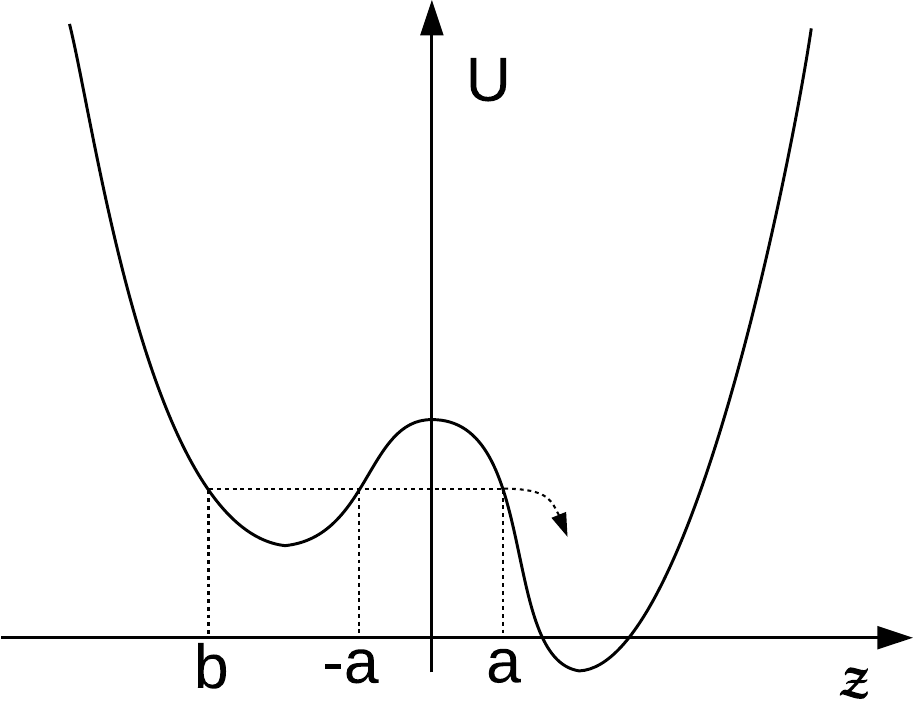}
 \caption{the inter well potential modified by the electric field} 
 \label{fig3}
\end{figure}


\begin{thebibliography}{99}
\bibitem{1} New perspectives in QHE, ed. S. Das Sarma and A. Pinczuk, Willey NY (1997)
\bibitem{2} S.M. Girvin ,A.H.MacDonald, New perspectives in QHE
\bibitem{3} Stefano Luin et all, Phys.Rev.Let.94,146804(2005)
\bibitem{4} S.V. Iordanski, D.S. Lyubshin, J.Phys. Condens. Matter 21,45601 (2009)
\bibitem{5} I.B.Spielman et all, Phys.Rev.Lett.v 84\#25, 5808,(2000)
\bibitem{6} S.V. Iordanski, Pisma v ZhETF vol. 99, iss. 9, 606 (2014)
\bibitem{7} L.D.Landau, E.M.Lifshits,Quantum mechanics,v.3,ch. 7, Moscow Fizmathlit (2002),NY Pergamon 
\bibitem{8} A.D.K.Finck et all, Phys.Rev.B 78, 075302(2008)
\bibitem{9} E.M.Lifshits, L.P.Pitaevski , Statistical Physics,part 2,Moscow Fizmathlit (2002),NY Pergamon    
\bibitem{10} C.Wang,T.Senthil, arXiv :1507.08290v2 cond.mat.str-el
\bibitem{11} G.Murthy,R.Shankar,arXiv:1508.06974 v2 cond.mat.str-el
\end{thebibliography}
\end{document}